# Evaluation of IoT-Based Computational Intelligence Tools for DNA Sequence Analysis in Bioinformatics

Zainab Alansari[1,2], Nor Badrul Anuar[1], Amirrudin Kamsin[1], Safeeullah Soomro[2] and Mohammad Riyaz Belgaum[2]

[1] College of Computer Science and Information Technology,
University of Malaya, Malaysia
z.alansari@siswa.um.edu.my, { badrul, amir }@um.edu.my
[2] College of Computer Studies,
AMA International University, Kingdom of Bahrain
{zeinab, s.soomro, bmdriyaz}@amaiu.edu.bh

**Abstract.** In contemporary age, Computational Intelligence (CI) performs an essential role in the interpretation of big biological data considering that it could provide all of the molecular biology and DNA sequencing computations. For this purpose, many researchers have attempted to implement different tools in this field and have competed aggressively. Hence, determining the best of them among the enormous number of available tools is not an easy task, selecting the one which accomplishes big data in the concise time and with no error can significantly improve the scientist's contribution in the bioinformatics field. This study uses different analysis and methods such as Fuzzy, Dempster-Shafer, Murphy and Entropy Shannon to provide the most significant and reliable evaluation of IoT-based computational intelligence tools for DNA sequence analysis. The outcomes of this study can be advantageous to the bioinformatics community, researchers and experts in big biological data.

## 1 Introduction

Internet of Things is a new revolution in the Internet. Objects make themselves recognizable and getting smarter by creating and providing relevant decisions. They can connect to each other and can have access to collected information by other objects or can be a part of a larger complex service. This development coincides with the emergence of cloud computing capabilities and the transition from the traditional Internet to the IPv6 unlimited addressing capacity.

Computational Intelligence (CI) is one of the most significant sectors of AI which applies a variety of methods for the AI realization. The tools used in computational intelligence are often mathematical tools that somehow inspired by nature and the world around. The following are some of the most valuable tools and templates that considered in computational intelligence:

- Evolutionary computation which is a set of methods that are known as evolutionary algorithms. The most popular algorithms are genetic algorithm inspired by the theory of evolution and genetics. This algorithm stimulated the evolution process that happened in nature over millions of years. The primary applications of evolutionary algorithms are solving optimization problems and mathematical planning [1].
- Swarm intelligence and the methods that fall into this category suggest another model for solving optimization problems. In this way, a significant number of very simple and low intelligence agents collaborate or compete to form a different type of swarm intelligence or collective intelligence. For example, an ant colony optimization algorithm which simulated by the collective presence of ants is one of the swarm intelligence algorithms [2].
- Artificial Neural Networks (ANN) is also one of the most important CI algorithms. Almost all scientists are confident that the human brain is the most known complex structure in the entire universe. Mathematicians and engineers of artificial intelligence that inspired by the findings of neuroscientists (neurologist) introduced an ANN which uses a variety of information modeling and classification. Perhaps neural networks can be considered as the most valuable tool in machine learning field [3].
- Fuzzy systems are using concepts like high or low to describe an idea instead of using exact numbers. For example, in phrases such as high profit, the amount of profit is not exactly clear [4]. Today, fuzzy systems excess usage to design different smart appliances. In addition to the above, other mathematical tools are used to improve the overall performance of systems based on computational intelligence. The primary goal of researchers in the fields of artificial intelligence and computational intelligence is to create such tools that provide us a closer alliance with human intelligence [5].

In this paper, we focus on the evaluation of IoT based computational intelligence tools for DNA Sequence analysis and the most important challenges and open issues.

## 2 Literature Review

Bioinformatics handles the immature data collected from researchers daily to form the image, charts, and numbers. It also sorts the data gathered from a large variety of databases on the network. The meaning and significance of initial tests on data collection including experimental errors, principles or to data collection for statistical coincidence means, needs careful experimental design and multiplicity results [6]. Experiments in professional conditions, reactants, equipment and time, are costly. To conclude, biological data are always incomplete [7].

Driven large amounts of data from recent biological tests led to the creation of massive databases that contain genes, proteins, and genetic data and another data type's sprocket. [8] introduced big data and reviewed related technologies, such as cloud computing, Internet of Things and Hadoop. Researchers insist on recovering data from some of the central databases specification such as nuclide or amino acid chain, organism, marginal genes or proteins name [9]. To increase the production of experimental data, Computer simulations based on CI play a fundamental part in biological methods. [10] Considered the IoT criteria's in the health sector for sustainable development. Quick Result and conclusion based on CI increased the biological information products. [11] Discussed the relationship between sequence database, IoT, and bioinformatics.

Genetic engineering refers to a set of methods which are used for isolation, purification and implying and expression of a particular gene in a host [12]. It ultimately causes a particular trait or produces the desired product in the host organism. Today, the technology and knowledge of genetic engineering and molecular biotechnology seem almost unlimited [13]. In recent years, development of tools for DNA sequencing provided recombinant revolutions in the treatment of many human diseases including all kind of cancers and most of the autoimmune diseases such as diabetes and the detection, prevention and treatment of many congenital diseases [14].

With the development of technology and biologically advanced tools, researchers faced the massive amounts of big biological data which the analysis of them using experimental methods associated with some challenges [15]. Therefore, some new ways with high speed and accuracy needed for this purpose. Due to the high speed and accuracy of computational intelligence methods, it can be said that they are a good alternative for being considered instead of laboratory procedures [16].

## 3 Research Methodology

This study used a practical descriptive survey given that it is based on the decision team to provide the needed data for determining the considerable sample size. The research's data collected from engineers, researchers, and experts by questionnaires and interviews.

### 3.1 Theory of Dempster-Shafer

In uncertainty time, data integration is imperative, and for this purpose, Bayesian theory, fuzzy logic, and the evidence theory are known methods. The theory of Dempster-Shafer considered as one of the most methods used for uncertainty reasoning, modeling, and accuracy of intelligent systems [17].

Dempster-Shafer theory is one of the primary methods for evaluating the uncertainty of unstructured data. It was founded by Dempster using the concept of upper and lower probabilities, and then Shaffer introduced it as a theory [18]. Moreover, measurement of uncertainty is one of the most important roles of entropy as one of a basic concept of big data and can be used as an uncertainty analysis tools in a particular situation.

The uncertainty decision is one of the most important research issues in computational systems and unstructured data. In recent years, the researchers and engineers provided useful definitions of uncertainty. The dual uncertainty nature is expressed by Helton [19] with the following definitions:

1. Aleatory uncertainty: as the fact that system can act randomly.

2. Epistemic uncertainty: happens when there is a shortage of data about the particular system, and it is a feature for performance analyzing.

Dempster combination rule is critical to combine evidence from different sources [20] and is a potential tool to evaluate the risk assessment and computational results. This method used in the impossibility of test's accurate measurement or inference knowledge of expert's opinion. One significant feature of this theory is the combination of evidence which extracted from different sources and modeling of conflict between them.

### 3.2 Reliability of Research Tools

According to the prepared questionnaires, it can be said that they measure all the criterions and options. In another word, most of the questions contain the desired structure considering that all the criterions investigated and the designer was not able to design an absolute orientation in questionnaire's design. Moreover, interviews were conducted to determine the security level of criterions and the overall rating level were calculated using the fuzzy and Dempster-Shafer method which they obtained outputs were logically correct. The measurement's reliability in this study doesn't benefit from the quantitative methods thus the assessment credit of assessors is to be considered as a criterion for reliability analysis. However, for paired comparison of the questionnaires which is based on the saaty's scale [21], we can use the compatibility rating to evaluate the reliability. Therefore, the compatibility criterion which is used to measure the incompatibility in the paired comparison matrix, when we use the group AHP to combine the matrixes, is determined as follow:

$$CI = \frac{(\lambda_{max}-n)}{n}.$$ (1)

Then Compatibility Rating or CR will be calculated by $= \frac{CI}{RI}$. If the result of CR is less than 0.1, the matrix compatibility is acceptable. The research process is shown in figure 1.

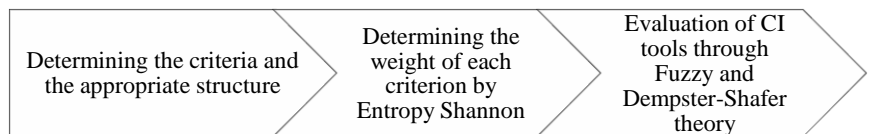

**Fig. 1.** The process of study.

### 3.3 Research Process

1) First Stage
At this stage of research, all the computation tools for DNA sequence analysis in bioinformatics identified through library research, literature review, and all current researches in this area. Then they were placed in a cycle form with an appropriate structure which a total of fourteen criteria constitute the main elements of the cycle. Finally, the obtained components and subcomponents were verified by bioinformatics experts. The criterion's structure is shown in figure 2.

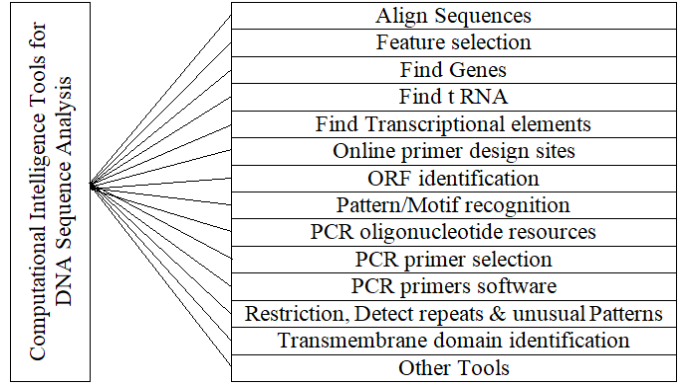

**Fig. 2.** Computational Intelligence tools for DNA sequence analysis.

2) Second Stage

The aim of this phase is to determine the importance of criterions using entropy analysis of Shannon [22]. Therefore, to achieve the best result, a questionnaire was given to some experts of bioinformatics and the specific matrices to each were formed. Due to the formation of multiple matrices, we should use the geometric mean for variable of each matrix $D = \left\| a_{ij} = \frac{w_i}{w_j} \right\|$ to obtain the final matrix. Equation 2 has been used to calculate the weights of each variable:

$$a'_{ij} = \left( \prod_{i=1}^{k} a_{ijI} \right)^{\frac{1}{k}}. \tag{2}$$

By getting the integrated matrix from equation 2, compatibility rating calculated which is used as an input for entropy Shannon. Finally, the final weight of each variable is computed using entropy Shannon noting that the input and output of each stage in this study are different from each other and each part is covering some specific requirements of this research. Therefore, each of fuzzy and entropy methods used in this study has different output. In this regard, to determine the variable's weight, the steps below are followed. The decision matrix contains some data which is used in entropy as an evaluation criterion. Suppose that the obtained decision matrix using paired comparison and combining geometric mean is as table 1.

**Table 1.** Decision about indicators.

| Indicator | $C_1$ | $C_2$ | … | $C_n$ |
|---|---|---|---|---|
| $C_1$ | $a_{11}$ | $a_{12}$ | … | $a_{1n}$ |
| $C_2$ | $a_{21}$ | $a_{22}$ | … | $a_{2n}$ |
| … | … | … | … | … |
| $C_n$ | $a_{m1}$ | $a_{m2}$ | … | $a_{mn}$ |
| $W_j$ | $W_1$ | $W_2$ | … | $W_n$ |

Using this matrix $P_{ij}$ is calculated as equation 3:

$$P_{ij} = \frac{a_{ij}}{\sum_{i=1}^{m} a_{ij}} \; ; \; \forall_{i,j}. \tag{3}$$

The Entropy's indicator $E_j$ is obtained by Equation 4:

$$E_j = -k \sum_{i=1}^{m} [P_{ij} \ln P_{ij}] \; ; \; \forall_j. \tag{4}$$

Uncertainty or deviation degree $d_j$ which obtained for indicator j, shows that the specific indicator of j, how much useful information provides for the decision. The amount of $W_j$ is obtained from equation 5:

$$d_j = 1 - E_j\ ;\ \forall_j\ . \qquad (5)$$

Then the weight is calculated using equation 6:

$$W_j = \frac{d_j}{\sum_{j=1}^{n} d_j}\ ;\ \forall_j\ . \qquad (6)$$

If a particular weight was considered earlier like $\lambda_j$ for indicator j, the adjusted weight of $W_j^{'}$ is calculated as equation 7:

$$W_j^{'} = \frac{\lambda_j W_j}{\sum_{j=1}^{n} \lambda_j W_j}\ ;\ \forall_j\ . \qquad (7)$$

3) Third Stage

In this phase, the assessment of CI tools for DNA sequence analysis should be evaluated. By collecting the considered data, the next level starts which is data analysis and evaluation of final decision [23]. In order to determine the level of each tool, the collected data from fourth questionnaires used as input in fuzzy functions. Primary motivations for the fuzzy sets is uncertainty. A characterized function can define each subset of fuzzy A in the main set of X. These functions are called the membership function which for each x member, from the central set X, allocate a number in the range of 0.1 which represents the degree of x membership in the fuzzy set of A. Therefore it is defined as A: X [1.0]. An example of a fuzzy set A in a defined set of X is $A = \{\langle x, \mu_A(x)\rangle | x \in X$ which $\mu_A: X \to [0.1]$ is the membership function of A. the real value of $\mu_A(x)$, describes the degree of $x \in X$ in A. for a finite set of A=$\{x_1, \ldots, x_i, \ldots, x_n\}$, the fuzzy set of (A, m), usually shown as $A = \left\{\frac{\mu_A(x_1)}{x_1}, \ldots, \frac{\mu_A(x_i)}{x_i}, \ldots, \frac{\mu_A(x_n)}{x_n}\right\}$.

In this study, if X is a defined set, five different variables describe the degree of CI tools in DNA Sequence Analysis which are X= {(VL) Very Low, (L) Low, (M) Medium, (H) High, (VH) Very High}. If we assume that only two adjacent variable overlaps, the fuzzy functions are defined as follows:

$$f_{very\ low}(x) = -0.4x + 1, \qquad 0 \leq x \leq 2.5\ . \qquad (8)$$
$$f_{low}(x) = -0.4x, \qquad 0 \leq x \leq 2.5\ .$$
$$f_{low}(x) = -0.4x + 2, \qquad 2.5 \leq x \leq 5\ .$$
$$f_{medium}(x) = 0.4x - 1, \qquad 2.5 \leq x \leq 5\ .$$
$$f_{medium}(x) = -0.4x + 3, \qquad 5 \leq x \leq 7.5\ .$$
$$f_{high}(x) = 0.4x - 2, \qquad 5 \leq x \leq 7.5\ .$$
$$f_{high}(x) = -0.4x + 4, \qquad 7.5 \leq x \leq 10\ .$$
$$f_{very\ high}(x) = 0.4x - 3, \qquad 7.5 \leq x \leq 10\ .$$

Which $f_{VL}$, $f_L$, $f_M$, $f_H$ and $f_{VH}$ are the membership functions of fuzzy sets. After determining the degree of each indicator, it is time to combine the same level functions. For this purpose, we must lower the functions to increase the confidence given to each indicator. In fact, the lower rate is used when an information source provides a Basic Probability Assignment ($BPAm$) which has same reliability as $\propto$. Therefore, (1-$\propto$) considers as a lowering rate and the new $BPAm^{\propto}$ is defined as:

$$m^{'}(A) = \propto m(A),\quad \forall A \subset \theta,\quad A \neq \theta\ . \qquad (9)$$
$$m^{'}(\theta) = 1 - \propto + \propto m(\theta)\ .$$

All the mass functions should be lowered using ∝ which is called the lower factor where m is a mass function of a witness, $m^a$ represents the indicative allocation function of initial lower probability and the lower factor a(0≤a≤ 1), determine the evidence reliability. Noting that before the final composition the overlap value of indicators must be calculated by equation 10:

$$m'(A) = \propto m(A), \quad \forall A \in \theta, \quad A \neq \theta . \tag{10}$$
$$m'(\{Y, A\}) = \frac{S(Y \cap A)}{S(X \cap A)} \times (1 - \propto m(A)), \quad Y \neq A, \quad Y \in X, \quad X \subset \theta .$$

Then the combination level began. Given that this study contains some conflicts, an averaging method of Murphy [24] is used to overcome the conflicts. As Murphy proposed, if all evidences are available concurrently, the mass average can be calculated and find the final mass by joining the averaged values several times. This rule can combine the two $BPA(m)$ of $m_1$ and $m_2$ for the new $BPA(m)$. Noting that the Dempster combination rule combines the multiple belief functions through $BPA(m)$. Dempster-Shafer combination rule is shown as $m = m_1 \oplus m_2$ and specifically obtain from the combination of BPAs $m_1$ and $m_2$:

$A \neq \emptyset$ and $m_{12}(\emptyset) = 0$ when (11)
$$m_{12}(A) = \frac{\sum_{B \cap C = A} m_1(B) m_2(C)}{1 - k} .$$
$$k = \sum_{B \cap C = \emptyset} m_1(B) m_2(C) .$$

## 4 Analysis and Result

According to the decision matrix of table 1, to gain the indicator's weight, we have to follow the steps below:

*Step one*; calculating $P_{ij}$: after calculating $P_{ij}$ and gaining its values we follow the other steps as below.

*Step two;* Calculate the entropy amount $E_j$: according to the calculated values of $P_{ij}$ and Equation 4, the amount of entropy can obtain which is shown in table 2.

**Table 2.** Gained values (step 2 to 5).

| Indicators | Entropy value ($E_j$) | Uncertainty value ($d_j$) | Indicator's weight (($W_j$) | Intellectual weight ($\lambda_j$)) | Adjusted weight ($W_j'$) |
|---|---|---|---|---|---|
| B1 | 0.966 | 0.034 | 0.245 | 0.2333 | 0.313 |
| B2 | 0.963 | 0.037 | 0.263 | 0.2333 | 0.336 |
| B3 | 0.985 | 0.015 | 0.106 | 0.2333 | 0.135 |
| B4 | 0.982 | 0.018 | 0.13 | 0.1834 | 0.13 |
| B5 | 0.977 | 0.023 | 0.166 | 0.0667 | 0.061 |
| B6 | 0.987 | 0.013 | 0.091 | 0.05 | 0.025 |
| B7 | 0.856 | 0.144 | 0.0293 | 0.4 | 0.355 |
| B8 | 0.735 | 0.265 | 0.54 | 0.3 | 0.492 |
| B9 | 0.918 | 0.082 | 0.167 | 0.3 | 0.152 |
| B10 | 0.996 | 0.004 | 0.024 | 0.1667 | 0.27 |
| B11 | 0.965 | 0.035 | 0.184 | 0.1333 | 0.17 |
| B12 | 0.975 | 0.025 | 0.13 | 0.15 | 0.136 |
| B13 | 0.992 | 0.008 | 0.043 | 0.15 | 0.045 |
| B14 | 0.945 | 0.055 | 0.129 | 0.1833 | 0.372 |

*Step three*; Calculating Uncertainty value ($d_j$): the values of uncertainty are gained according to entropy's values and equation 5.

*Step four;* calculating the weight ($W_j$): the weight of each indicator is gained according to the uncertainty value and equation 6.

*Step five*; Adjusted weight ($W_j'$): the adjusted weights are calculating according to indicator's weight and Intellectual weight ($\lambda_j$) is calculating according to equation 7.

Based on table 2 and calculating the mean of entropy and uncertainty values and the weight of indicators with intellectual and adjusted weight, we find that almost all fourteen indicators have near rating and similarities which are shown in figure 3.

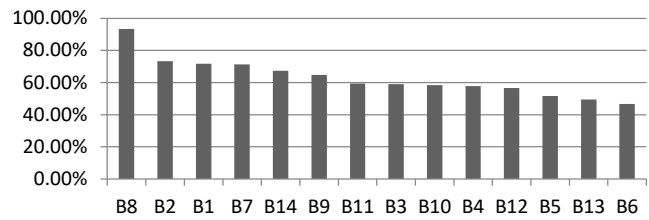

**Fig. 3.** Evaluation of IoT-Based CI Tools for DNA Sequence Analysis.

After calculating each indicator's weight, the second part of data which are usability, reliability, validity, and power of each indicator were collected from the experts and used as input for equation 8. Table 3 shows the calculated results:

**Table 3.** Indicator's rating.

| Indicators | Description | Rating |
|---|---|---|
| B1 | Align Sequences | M |
| B2 | Feature selection | VH |
| B3 | Find Genes | H |
| B4 | Find t RNA | M |
| B5 | Find Transcriptional elements | M |
| B6 | Online primer design sites | VH |
| B7 | ORF identification | H |
| B8 | Pattern/Motif recognition | H |
| B9 | PCR oligonucleotide resources | H |
| B10 | PCR primer selection | VL |
| B11 | PCR primers software | VL |
| B12 | Restriction, Detect repeats & unusual Patterns | L |
| B13 | Transmembrane domain Identification | VL |
| B14 | Other Tools | H |
| B14 | Other Tools | H |

By determining the indicator's score, the next step is to combine the indicators of each group. For this purpose, the following five diagnosis hypotheses considered: $\theta=$ {(VL)Very Low, (L)Low, (M)Medium, (H)High, (VH)Very High}

Each one of these is indicating the CI tools rating for DNA Sequence Analysis in Bioinformatics and used as input in Dempster-Shafer Theory. Noticing this evidence is preliminary and vague for combination, they need to lower first. Equation nine is used for lowering the evidence, and the overlap between variables obtained using equation 10 and finally is composition stage turn. After synthesizing the evidence, maybe 100% assurance allocate to the particular focal element. Several ways have been introduced for facing such conflicts. This study used Murphy's proposed idea, and the calculations results are shown in table 4.

**Table 4.** Overall evaluation of IoT-Based CI Tools for DNA Sequence Analysis.

| Evidence Combination | VL | L | M | H | VH | VL, L | L, M | M, H | H, VH |
|---|---|---|---|---|---|---|---|---|---|
| B1, B2, B3, B4 | 0 | 0 | 0 | 0 | 0 | 0 | 0.01 | 0.02 | 0.03 |
| B3, B4, B5, B6 | 0 | 0 | 0 | 0 | 0 | 0 | 0.01 | 0.02 | 0.01 |
| B5, B6, B7, B8 | 0 | 0 | 0 | 0.2 | 0 | 0 | 0.01 | 0.11 | 0.11 |
| B7, B8, B9, B10 | 0 | 0 | 0 | 0.2 | 0 | 0 | 0 | 0.08 | 0.08 |
| B9, B10, B11, B12 | 0 | 0 | 0 | 0.1 | 0 | 0.02 | 0.01 | 0.04 | 0.04 |
| B11, B12, B13, B14 | 0 | 0 | 0 | 0.1 | 0 | 0.02 | 0.01 | 0.03 | 0.03 |
| Average | 0 | 0 | 0 | 0.1 | 0 | 0.01 | 0.01 | 0.05 | 0.05 |

## 5  Conclusion

As per the results of this study, it indicates that pattern/motif recognition tools have the highest ranking and the lowest ranking is given to online primer design sites. This study shows that the usability of all the DNS sequence analysis methods integrated with the computational intelligence tools are almost equal and it confirms the importance of CI in bioinformatics. This study uses distinctive analysis and approaches such as fuzzy system, Dempster-Shafer algorithm, a method of Murphy and Shannon's entropy to provide the most significant and reliable evaluation of computational intelligence tools for DNA sequence analysis. The CI tools play a fundamental role in bioinformatics DNA sequence analysis. Hence, the recommendation to the computer scientists, engineers and researchers are to examine research, produce and propose innovative tools and methods in this area using the presented consequences of this research. The findings of this study can be advantageous to the bioinformatics community, researchers and experts in big biological data.